# UV Fluorescence Sensing of Medical-Chemical-and- Biological Analytes by an AlGaN/Si-on-Sapphire Optoelectronic Chip

Richard Soref

*Richard Soref is with the Engineering Department, University of Massachusetts at Boston, Boston, MA 02125 USA*

***Abstract-*** **-This paper proposes and analyzes a compact fluorescence-photometer lab-on-a-chip that is simpler than a fluorescence spectrometer, and that does not require spectrum reconstruction by a computer. An on-chip AlGaN VCSEL lasing at 280 nm, or at 310 nm, pumps an on-chip sample in the form of liquid or solid or tissue, thereby exciting "signature fluorescence" in the 300 to 540 nm wavelength range. This emission signal is detected and read out by an on-chip UV-enhanced Silicon photodiode. The 2 cm x 5 cm chip consists of C-plane (0001) Sapphire, and no lenses, prisms, or mirrors are utilized in the photometer. The chip is opto-electronic and features co-packaged electronics using tiny electronic chiplets that are bonded to the sapphire. These consist of an LD driver and transimpedance amplifier for the UV photonics, together with ASICs, ADC, microcontroller, micro-battery and RF-wireless or Ethernet chiplets. The "mission" of this self-contained cost-effective chip is sensing targeted medical-chemical-biological (MCB) agents in a local** **or network setting. Eight significant MCB applications are detailed here.**

*Index Terms*—**Ultraviolet photonics, med-chem-bio sensing, fluorescence photometry, AlGaN/sapphire photonics, optoelectronic integration.**

## I. INTRODUCTION

A fluorescence spectrometer is a high-resolution instrument that measures the detailed shapes-and-curves of a sample's fluorescent-emission spectrum; whereas, a fluorescence photometer is an instrument that measures the intensity (or the absence) of a sample's fluorescence within a broad spectral band. Such a photometer can be quite valuable, as detailed here.

The present paper proposes and analyzes a "fluorescence-photometer chip" that enables a large number of significant UV applications. An important aspect of this chip is that it measures analytes within a liquid or solid or skin. Traditionally, the word "chip" refers to a substrate about 5 mm x 10 mm in size. Such chips are often mounted on ceramic interposers that carry some electronic circuits. In the present paper, the chip substrate area is expanded to 2 cm x 5 cm. Here, this enlarged chip provides-free space UV photonics, and it also serves as a substrate for associated electronic ICs-and-battery. In other words, the chip provides both interposer and sensor functions —it is optoelectronic and "complete." The chip here is intrinsic crystalline C- plane sapphire with (0001) orientation.

The design study of this paper begins with a detailed proposal of eight key-and-innovative applications of the chip. In all cases,

the chip offers a major advantage over the existing, present fluorescence-measuring instruments that are generally large, expensive, and laboratory-bound. The present approach is compact, simple, portable and cost-effective.

## II. IMPORTANT NEW APPLICATIONS

Eight new med-chem-bio sensing applications are proposed here. In each case, the UV wavelength used during laser excitation is listed, the physical state of the analytes is given, and the experimentally determined fluorescence spectra---taken from the literature -- are presented. The new opportunities for future development are detailed.

### *A.Skin-Sensor of AGE/Diabetes*

By contacting the fluoro chip to human skin, a non-invasive label-free sensor is created to detect collagen crosslinks and advanced glycation end-products (AGE) fluorophores embedded in the dermis. Here the laser excitation will be at 310 nm, and the resulting emission spectra [1, 2] are shown in Fig. 1(a). Because AGE auto fluorescence is linked to diabetes complications, cardio-vascular events and chronic kidney disease, the opportunity here is for early risk detection and monitoring. The chip's collective identification of analytes is fully adequate here because total detected intensity is the biomarker.

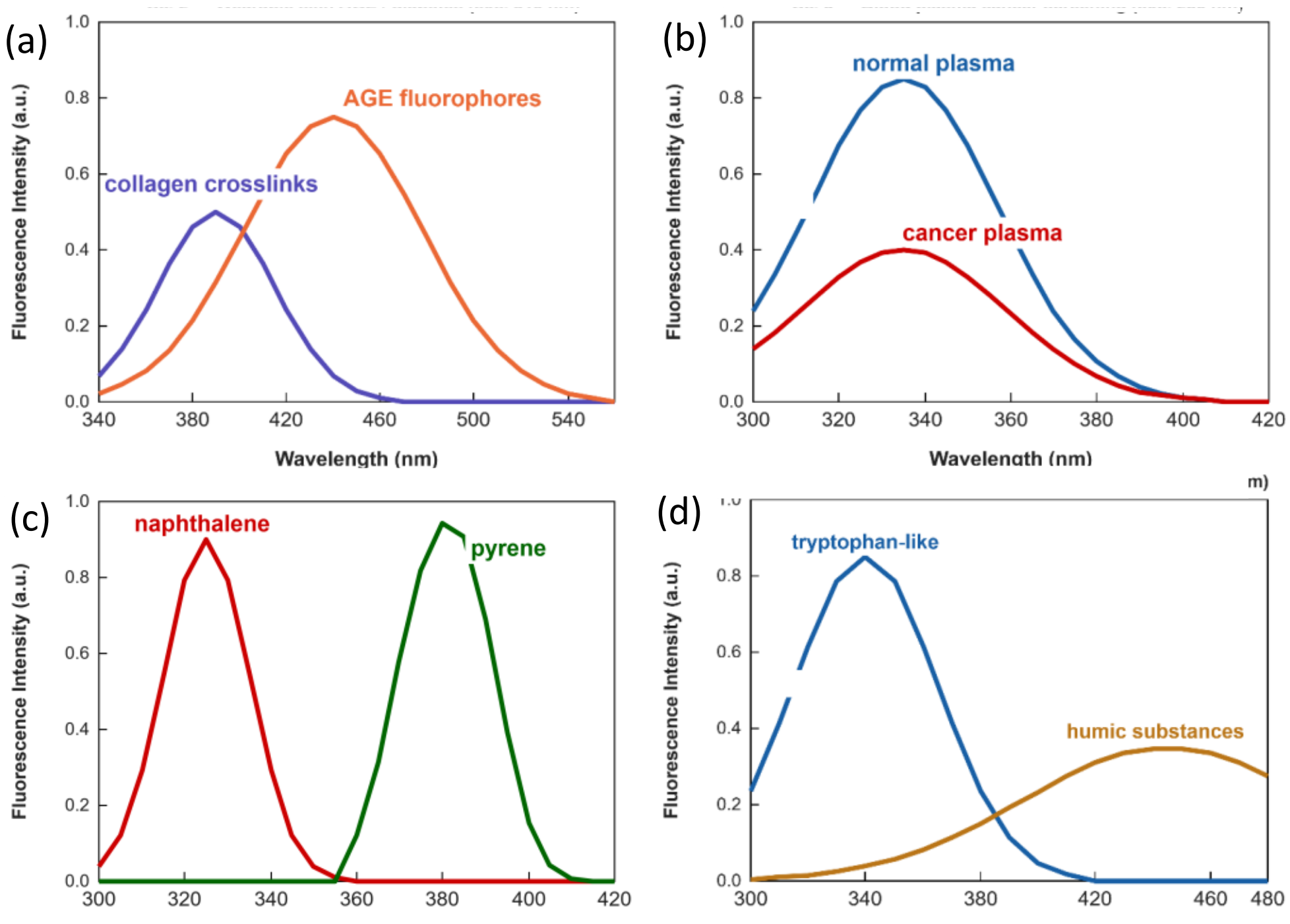

Fig. 1. Fluorescent-emission spectra of: (a) collagen crosslinks and AGE fluorophores in skin, (b) tryptophan in blood plasma, (c) naphthalene in and pyrene in water, and (d) bacterial cells in water.

## *B.Blood-Plasma Cancer Screening*

Tryptophan emission from blood plasma has diagnostic value. If the aqueous plasma is excited by 280 nm, the emission intensity in Fig. 1(b) [3, 4] can be used for cancer pre-screening of patients. The opportunity is to use a simple finger-prick blood draw for simple, low-cost, chip-based point of care. The chip's collective identification of analytes is fully adequate here because the change in intensity is the key metric, and a measurement of spectral shift is not required.

## *C.PAH Contamination in Water*

UV fluorescence, excited by a 275-nm laser, has become a quantitative tool for detecting polycyclic aromatic hydrocarbons (PAHs) in water. We select here the 280-nm excitation with the resulting spectra of naphthalene and pyrene dissolved in water [5, 6] shown in Fig 1(c). The 280-nm choice gives only a small reduction ~10% in naphthalene absorption. The opportunity here is networked PAH/oil-in-water sensors in river systems, drinking-water treatment plants, and oil-pipeline leak detection. The chip's collective identification of analytes is fully adequate here because the field question is the collective contamination presence.

*D. Water Quality—Bacterial*

A sample of water can contain dissolved humic substances and particulate suspended bacteria cells. For both of these analytes, 280-nm excitation will produce tryptophan-like fluorescence (TLF) from bacteria and broadband humic-substance emission as shown in Fig. 1(d) [7, 8]. The opportunity is to use the fluoro-chip as an acute-contamination event detector with a concentration-detection threshold of 30 to 50 parts-per-billion via TLF. A separate TLF measurement on a clean-water sample provides a baseline or reference for comparison with the "unknown-water" measurement. In that way, the chip is "self-referencing." Using a $TiO_2/SiO_2$ multi-layer interference $\lambda < 380$-nm short-pass filter on the photodetector (PD) face, it is possible to prevent the 380-nm to 500-nm portion of the humic fluorescence from reaching the PD. However, the TLF signal is likely to be much larger than the humic signal; so such filtering may not be necessary,

*E. Agricultural Plant Health*

Using 310-nm excitation of agricultural plant leaves, the resulting emission spectra of Fig. 2(a) [9, 10] can reveal the presence of ferulic acid and chlorogenic acid residing in plant cell walls and intercellular spaces. The opportunity here is to gauge plant health. The chip's collective identification of analytes is fully adequate here because the total phenolic index is the established metric.

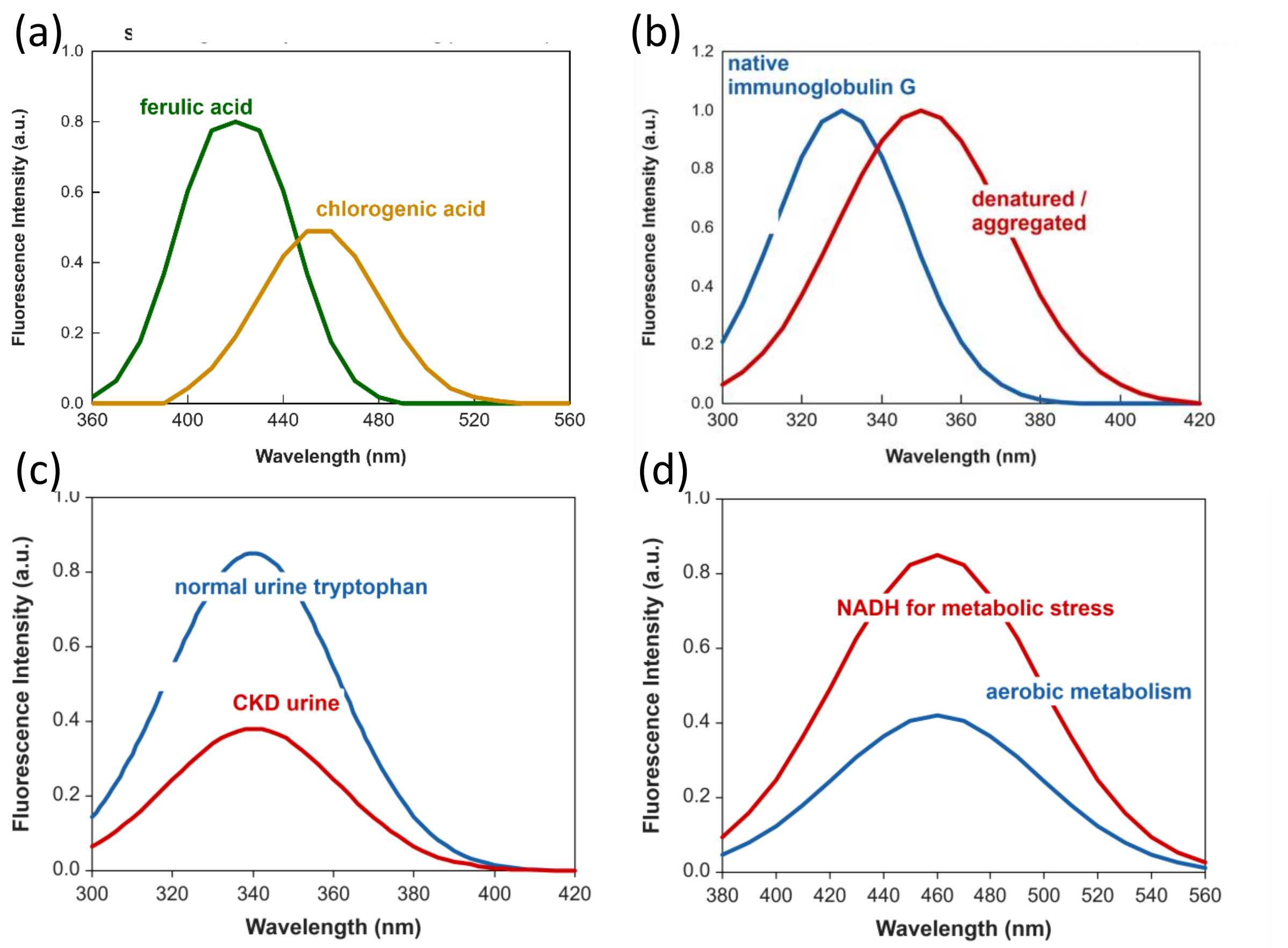

Fig. 2. Fluorescent-emission spectra of: (a) phenolic acids in plant leaves, (b) IgG in biopharma liquid, (c) excess Trp in urine, and (d) NADH in transplanted organs.

*F.Biopharma Quality Control*

Within the application-area of biologics manufacturing, real-time monitoring of monoclonal antibodies in bioreactor production is important, and tryptophan fluorescence plays a key role. The sample here is Immunoglobulin G within an aqueous buffer at 1–10 mg/mL concentration. The clear solution is fluorescence-measured in a flow-through cell bonded to the chip's top face. No sample preparation is needed. When the liquid is excited by the 280-nm laser, the resulting spectra shown in Fig. 2(b) [11,12] report both protein concentration and protein conformation. The opportunity here is a real-time process alarm--indicating protein loss, denaturation or aggregation. The alarm is triggered by a drop in fluorescence below a calibrated threshold. This Process Analytical Technology (PAT)-compliant measurement requires no sampling, no reagents, and no moving parts. Collective identification here of the total protein fluorescence state is sufficient because the use-case is threshold alarming, not species identification.

*G. Kidney Disease Monitoring by Urinary Trp Fluoro*

The concentration of Tryptophan (Trp) in urine is a key biomarker of Chronic Kidney Disease (CKD). The key urine spectra of Fig. 2(c) [13,14] show the fluoro signal strengths for normal and diseased persons. The opportunity here is early-stage disease detection using a simple, non-invasive test. The low-cost chip connects by WiFi to the patient's nephrology team. The sample sits on a small well on the chip's top surface and the sample requires no preparation.

Spectral-resolution sensing is not required for the Fig.-2(c) application because the disease biomarker is a decline in total Trp over time. There are two application scenarios and self-referencing of the chip is used in both.

Scenario-1: Longitudinal monitoring of a known CKD patient. Here the first step is to take a small, sealed vial of factory-calibrated phosphate-buffered saline containing 100 μM of tryptophan. The vial is placed on the chip well, and a fluoro-voltage V(ref) is recorded which corrects for vertical-cavity surface-emitting laser (VCSEL) power, PD responsivity and optical coupling. Next, the patient's fresh urine is placed in the chip well, and a voltage V(sample) is recorded. The ratio V(sample)/V(ref) is expressed in units of μM. At each clinic visit—say every two months—the patient provides a urine sample, and the calibrated chip will detect any decline in Trp.

Scenario-2: Population screening for kidney health. Here, a new patient with no prior baseline can learn whether his Trp level is abnormally low compared with that of a healthy population. Here the measured V(sample)/V(ref) ratio is compared with that of healthy adults, 100-400 μM Trp depending on dietary protein intake and time of day. A reading below, or well below this, would flag CKD.

### *H. Organ-Transplant Monitoring by NADH Fluoro*

An organ-perfusion machine is a portable medical device that keeps donated organs alive and functioning outside the body by continuously pumping blood or oxygen or a specialized solution through them. If an organ becomes deprived of oxygen during this process, mitochondrial nicotinamide adenine dinucleotide hydride (NADH) will accumulate in the organ. If the fluoro lab chip is pressed gently against the exposed surface of the donated heart or kidney or liver, then the laser pump will excite the outermost 300 um of this tissue, and the resulting fluoro signal, Fig. 2(d) [15,16] will reveal the NADH concentration in the organ, where NADH is related to organ metabolism. The opportunity here is real-time diagnosis of organ health. The fluoro signal is measured a few times in a time sequence. A rising NADH signal indicates worsening ischaemic injury. A steady or falling signal indicates that the organ is recovering and metabolising normally, which is a good prognostic sign for graft function after transplantation.

## III. PROPOSED CHIP ARCHITECTURE

The chip here does not have any photonic waveguides on it. Instead, it is a free-space chip. Also, the high transparency of sapphire in the UV-and-visible plays a major role in chip design. Specifically, the sample cell is mounted on the top face of the chip, as in Fig. 3, while all other components are mounted on the bottom face. Then there is two-way UV-vis "communication" within the $Al_2O_3$ chip body where pumping light and sample-emitted light travel between the bottom-and-top chip faces. Thereby, the use of lenses is entirely avoided. Before describing the chip layout, let us examine optimum choice of excitation laser, and the optimum choice of the photodetector.

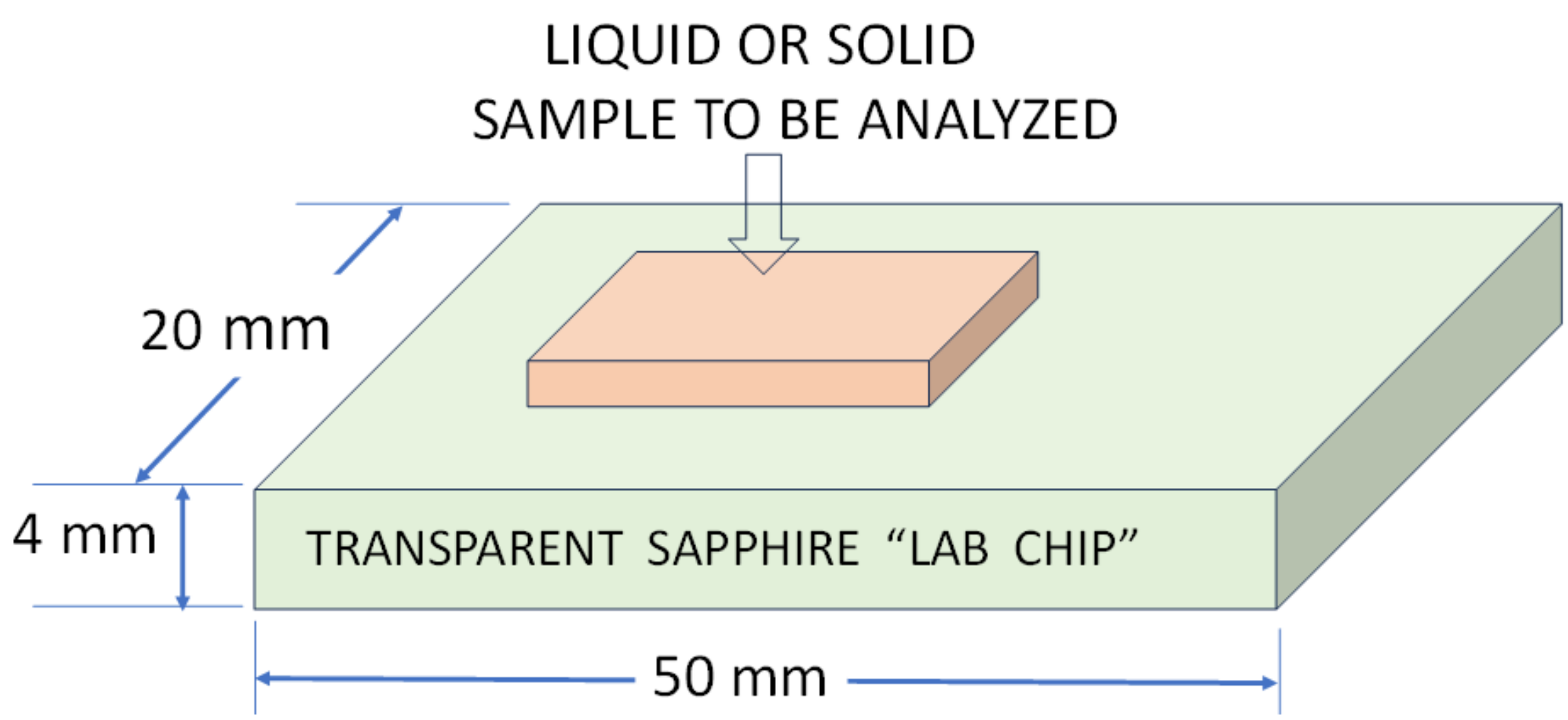

Fig. 3. Schematic top portion of fluorescence-photometer laboratory-chip.

### *A.AlGaN VCSEL Ultraviolet Pump*

In this photometer, there are only two electrically addressed photonic components: the laser diode and the detector diode. As described below, both of these components would be fabricated on a donor wafer and would be then transfer-printed/bonded to the bottom face of the sapphire chip. Of these two, the design of the laser presents the biggest challenge because several issues have to be confronted, beginning with the choice of an edge-emitting or surface emitting laser. Within the surface and edge categories, consideration must be given to the bonding-to-sapphire techniques, the optimum emitting area of the laser, the wire bonding methods, the need for multiple quantum wells (MQWs) in the active zone, and the cone angle of laser emission.

An examination of the literature [17-20] leads to the conclusion here that the VCSEL is the best choice. Also, the experimental reports on AlGaN VCSELs show that distributed Bragg reflectors (DBRs) comprised of multi-layer dielectrics-or-semiconductors, will do a good job of defining the laser cavity, and that MQWs produce a strong and collimated output [21,22]. Before discussing the VCSEL's area and die bonding, let us quantify the AlGaN compositions required.

Six of the sensing scenarios described above require a UV VCSEL emitting at an excitation wavelength of 280 nm, whereas two applications need 310 nm excitation. These apps correspond to photon energies of 4.43 eV and 4.00 eV. The required MQW VCSELs are quite similar, and differ mainly in the Al mole fraction of their AlGaN quantum wells and barriers. We now derive the required compositions.

Regarding the bandgap of $Al_{(x)}Ga_{(1-x)}N$, it follows the relation

$$E_g(x) = x\,E_g(\mathrm{AlN}) + (1-x)\,E_g(\mathrm{GaN}) - b\,x(1-x)$$

with $E_g$(AlN)=6.20 eV, $E_g$(GaN)=3.40 eV, and with the experimentally established bowing parameter $b$ =1.30 eV [23] giving

$$E_g(x) = 3.40 + 1.50\,x + 1.30\,x^2 \text{ (eV)}.$$

Regarding the quantum-confinement correction, the VCSEL gain medium consists of two to four AlGaN quantum wells, each 2–to-4 nm thick. Quantum confinement blueshifts the ground-state electron–heavy-hole transition above the bulk bandgap by approximately 100–200 meV; we adopt $\Delta E_{\mathrm{conf}} \approx$150 meV as representative of published AlGaN MQW devices in this thickness range [18, 20]. The VCSEL lases at the energy of this ground-state transition, so this photon energy is

$$h\nu_{\mathrm{VCSEL}} = E_g(x_{\mathrm{well}}) + \Delta E_{\mathrm{conf}}$$

where $x_{\mathrm{well}}$ is the Al fraction of the well material. Rearranging to find the required bulk bandgap of the well:

$$E_g(x_{\mathrm{well}}) = h\nu_{\mathrm{VCSEL}} - \Delta E_{\mathrm{conf}} = h\nu_{\mathrm{VCSEL}} - 0.150 \text{ eV}.$$

To obtain the well compositions, we substitute the two target photon energies. Solving for $x_{\mathrm{well}}$, we find for λ = 280 nm, $E_g$(well) = 4.28 eV and QW = Al(0.42)Ga(0.58)N; while for λ = 310 nm, $E_g$(well) = 3.85 eV and QW = Al(0.24)Ga(0.76)N.

Regarding barrier compositions, the barrier Al fraction is set approximately 0.15–0.20 higher than the well fraction to provide a conduction-band offset of ≳200 meV, sufficient to confine injected electrons within the well at room temperature. The resulting barrier compositions are: Al(0.60)Ga(0.40)N at 280 nm and Al(0.42)Ga(0.58)N at 310 nm. The 280-nm analysis here assumes that electrically injected AlGaN VCSELs at 280 nm, currently under active development will become available in the near term.

The emitting top face of the VCSEL is bonded to the chip's bottom face, while two electrical contacts are made to the exposed lower regions of the VCSEL. The bottom views of the chip, Figs. 4 and 5, illustrate the geometry chosen for the *p-i-n* laser diode.

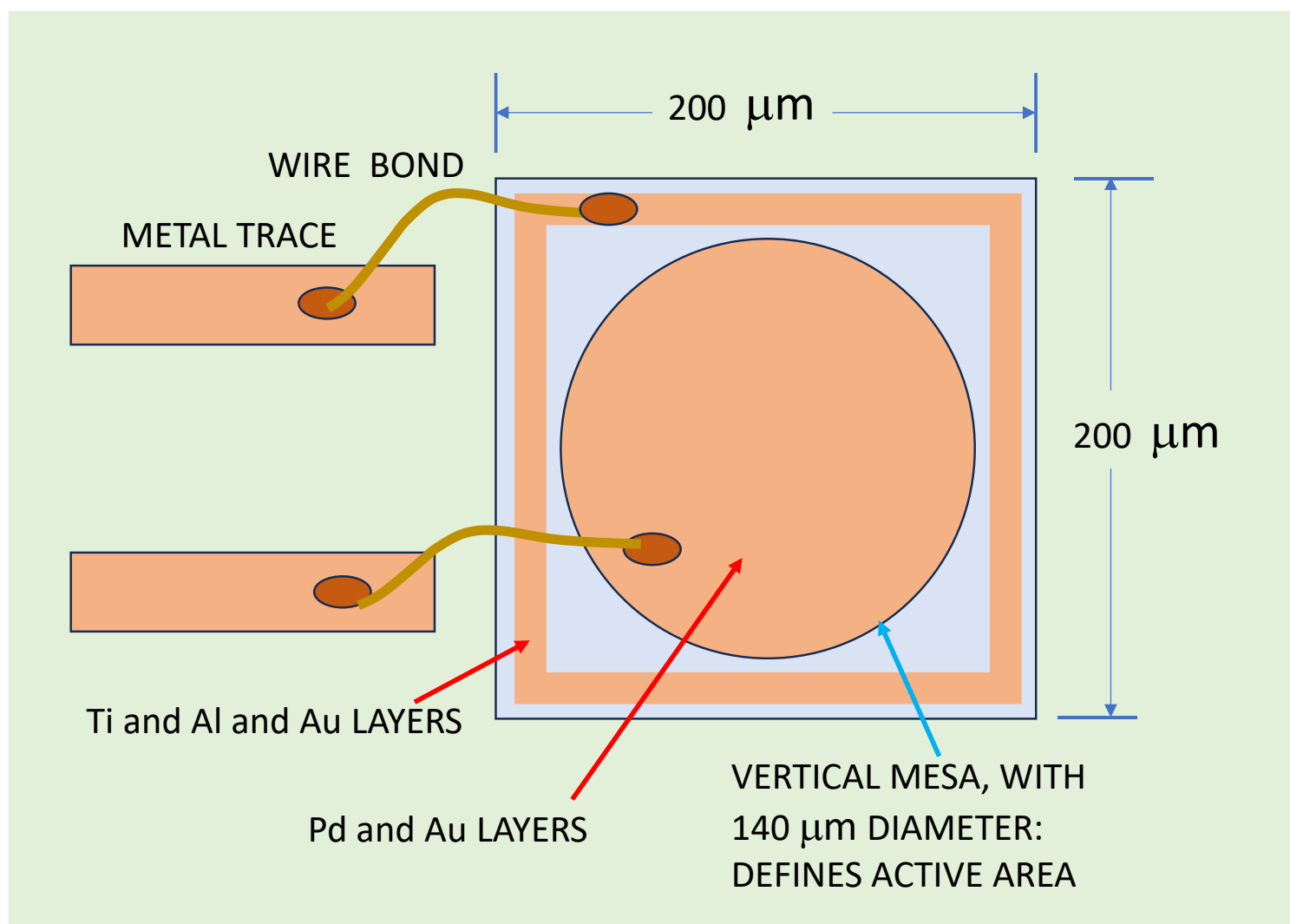

Fig. 4. Chip bottom view of VCSEL showing active mesa, *n*- and *p*-contacts and wire bonds.

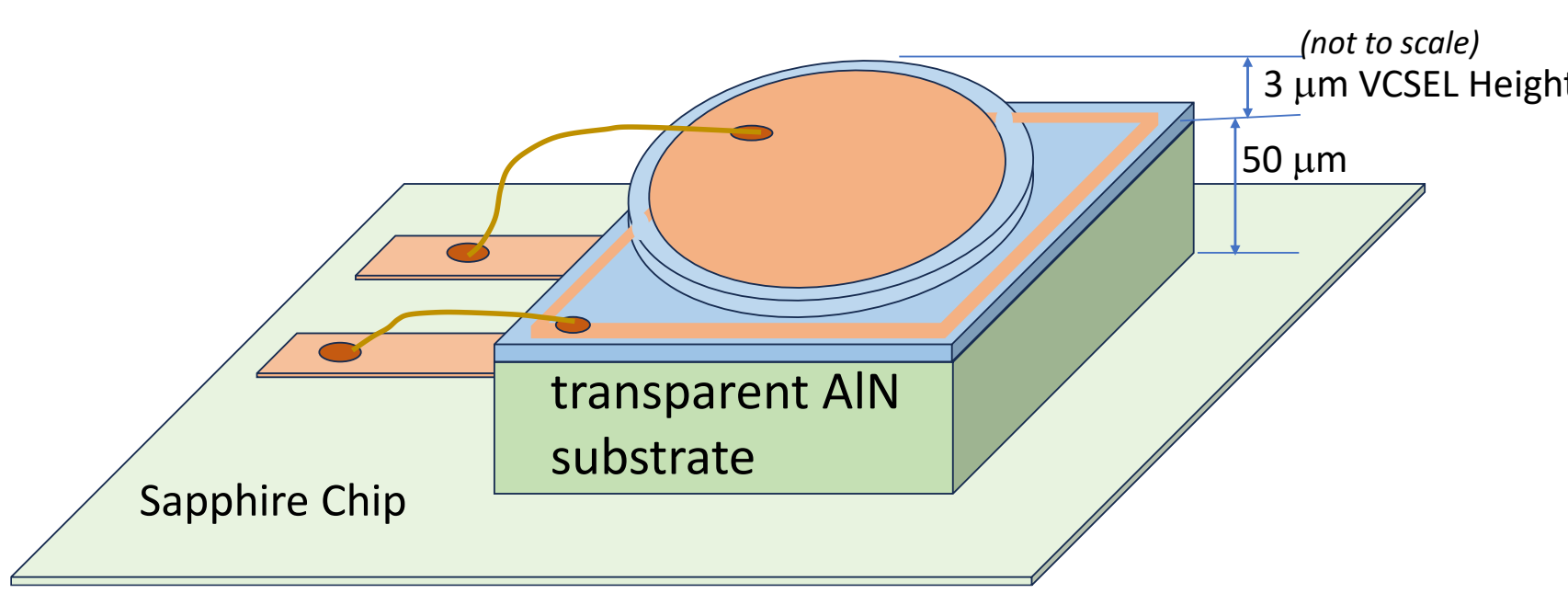

Fig. 5. Perspective of VCSEL bonded on chip bottom showing integral 50-um AlN laser-die substrate.

The laser consists of a *p*-type DBR/i-MQW "mesa" upon an *n*-type DBR rectangular frame. In order to attain the desired ~ 1 mW of laser output power, the active mesa is chosen to have 140 μm diameter, with 200 μm x 200 μm chosen for the rectangle size. The metal disc, the *p* pad contacting the p-DBR, consists of 5-nm Pd and 100-nm "reflective" Au, and a rectangular Ti/Al/Au *n*-pad contact is used as shown. Gold wires bonded to these two pads are then bonded to their respective metal traces on the chip surface.

The overall thickness of this VCSEL is estimated as 3 μm, which shows that the laser is a thin "film" device. This film is mechanically very fragile. If this film were to be lifted off of its donor wafer (for example by the liftoff methods given in the literature), it would be very difficult to transfer and bond the "film die" without breaking it. Thus, an alternative technique is needed. Here we propose a practical, mechanically sound approach in which the VCSEL is epitaxially grown upon a thick, intrinsic, crystalline AlN substrate, where the "thick" AlN becomes an integral part of the VCSEL "die." First, a commercial single-crystal AlN substrate would be purchased. Next, that substrate would be thinned to 50 μm by grinding and chemical-mechanical polishing, and after that the VCSEL would be grown epitaxially on that substrate. The resulting VCSEL die is illustrated in Fig. 5 which shows also its electrical contacts and leads as per Fig. 4.

Now let us go into more detail about the VCSEL layering. Figure 6 shows a cross-section view of the to-chip-bonded device, which is not-to-scale in the vertical direction in order see more clearly the layered construction. Theory indicates that 99.5% reflectivity will be achieved by the 13-period *p*-AlGaN DBR, with 99.0% reflectivity for the 15-period *n*-AlGaN DBR. The MQW construction has been detailed above in this section. The VCSEL literature indicates that multimode lasing will result from the 140-µm-diameter active region.

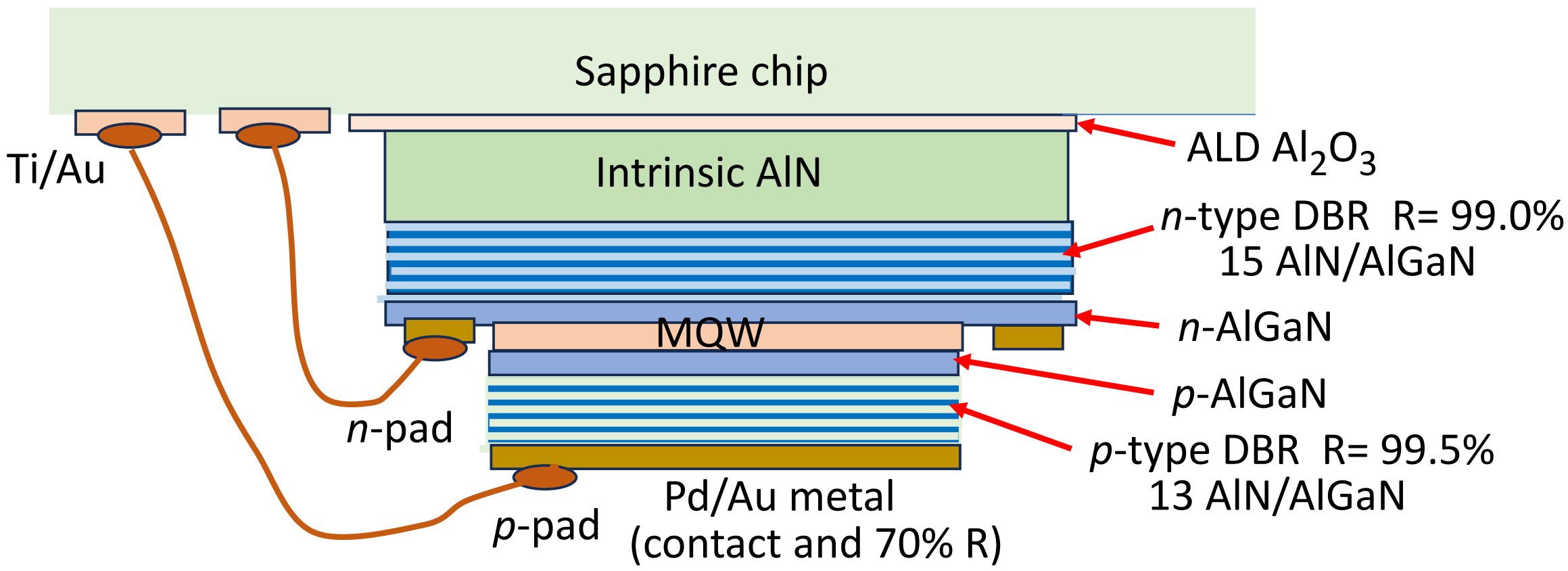

Fig. 6. Cross-sectional side view of VCSEL bonded to chip bottom.

Figure 7 illustrates the conical beam of laser emission anticipated from the Fig.-6 VCSEL. Using laser theory, a cone angle of 3.7 degrees FWHM is predicted for the Fig. 6-device. After this cone traverses the 4-mm chip body, it produces a laser-spot diameter of 0.4 mm at the sample's front surface.

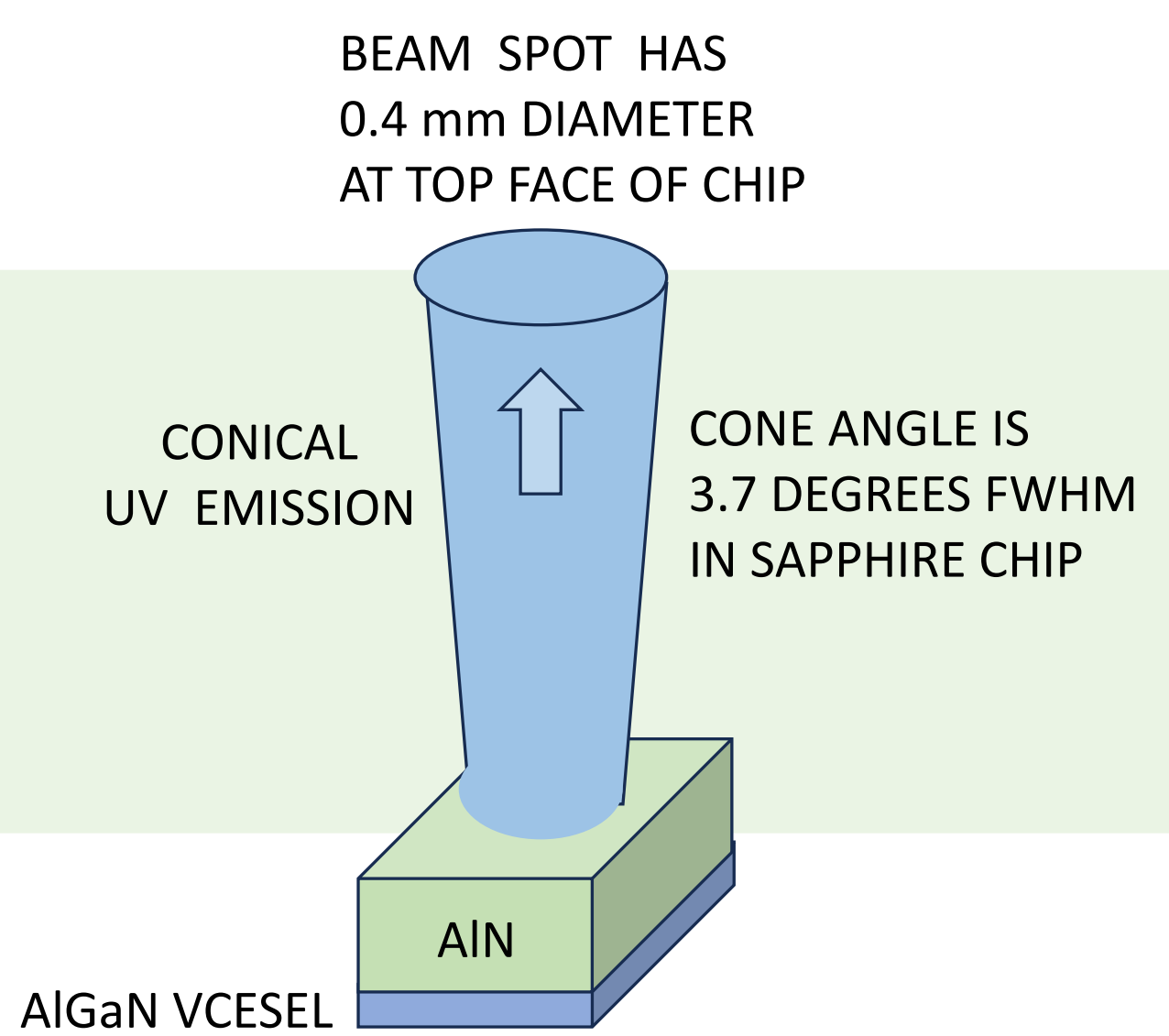

Fig. 7. Illustrating the narrow Cone of the laser beam traversing the 4-mm transparent chip.

### *B.UV-Enhanced Silicon Photodetector*

The chip uses only one PD. According to Figs. 1 and 2, t is desirable to have a PD that detects 280-to-560 nm adequately. The large-area Si PD is a fine choice for this assignment, however it requires UV enhancement [24,25] because UV-generated carriers are created in a very shallow layer at the entry face of the Si PD. Manufacturers have made practical progress in this area using shallow, low-defect *pn* or *pin* junctions with shallow carrier collection and with surface-process control. Today, UV-enhanced SiPDs are available from several companies including Hamamatsu, Osi Optoelectronics, Opto Diode Corp, Advanced Photonix, and Marktech Optoelectronics. We are not advocating any particular product and the PD specifications listed here are presented only as "place holders" that represent the practical performance that is readily attained.

An example is the S1336 33BQ product of Hamamatsu whose responsivity is shown in Fig. 8 where the spectral coverage is sufficient to read out the fluoro intensity in all eight of the applications listed above. The 33BQ detector is particularly relevant because its 2.4 mm x 2.4 mm sensing area is quite compatible with the "through-chip" sensing geometry shown in Section D below: that is, the PD captures a useful fraction of the fluorescence from the pumped Figs. 1-and-2 samples. The PD's 65 pF terminal capacitance, 30 pA dark current and low TIA noise floor improve the SNR.

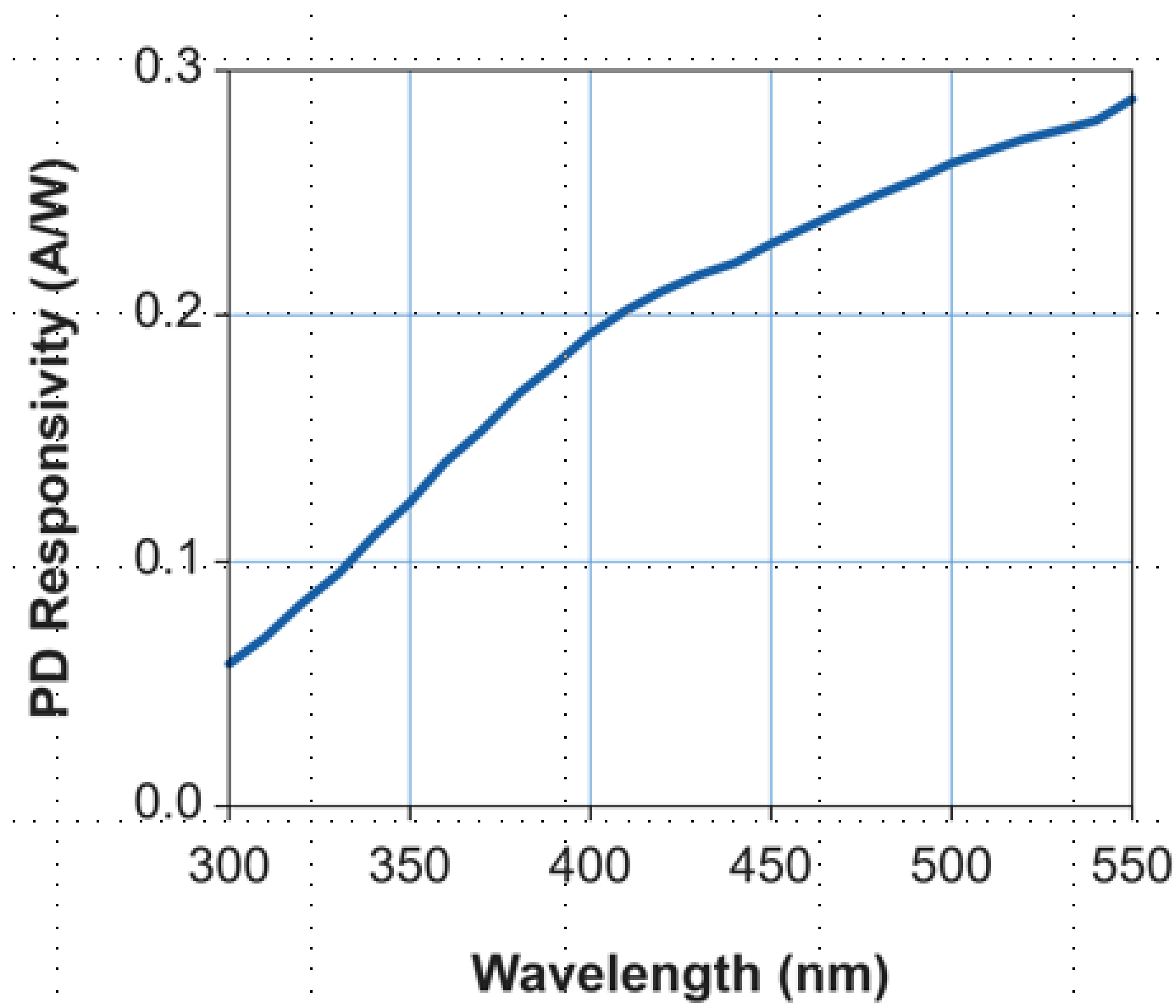

Fig. 8. Experimentally determined UV-Vis responsivity of a commercial UV-enhanced Si *p-i-n* photodiode.

*C .Laser-blocking Filter for Photodetector*

As shown below in Section D, the PD is positioned to capture an adequate portion of the total fluoro signal; however some scattered laser light will unavoidably reach the input face of the PD, and such light will obscure and “pollute” the signal. Thus, to prevent scattered laser-light from impinging upon the PD, it is important to create a laser-blocking long-pass filter that is placed in front of the PD. A good choice for this assignment is an AlGaN layer whose bandgap energy is slightly lower than the laser photon energy. We examined the UV absorption spectra of AlGaN and noted that the transition edge of an AlGaN filter is dramatically sharper than that of a dielectric filter, reinforcing the AlGaN choice.

The next step is to design a bulk absorptive (AlGaN bandgap) filter for 280-nm laser blocking as well as a second filter for 310-nm laser blocking. For the first case, we selected a bandgap wavelength of 313 nm; and for the second case, a bandgap wavelength of 334 nm was chosen. To calculate the absorption coefficient of an AlGaN filter, we used a parabolic-band model for $h\nu > E_g$, together with an Urbach exponential-tail model when $h\nu < E_g$. The above-cited 313-nm and 334-nm gap wavelengths lead to filter compositions of Al(0.30)Ga(0.70)N for the 280 nm laser, (filter A) and Al(0.18)Ga(0.82)N for the 310 nm laser. (filter B). Taking a 1.5 μm layer of Al(0.30)Ga(0.70)N for filter A and a 4.0 μm layer of Al(0.18)Ga(0.82)N for filter B, gives the transmission-versus-wavelength curves shown in Fig. 9. One option is to grow this AlGaN layer directly on the bottom of the sapphire chip in the PD-area location.

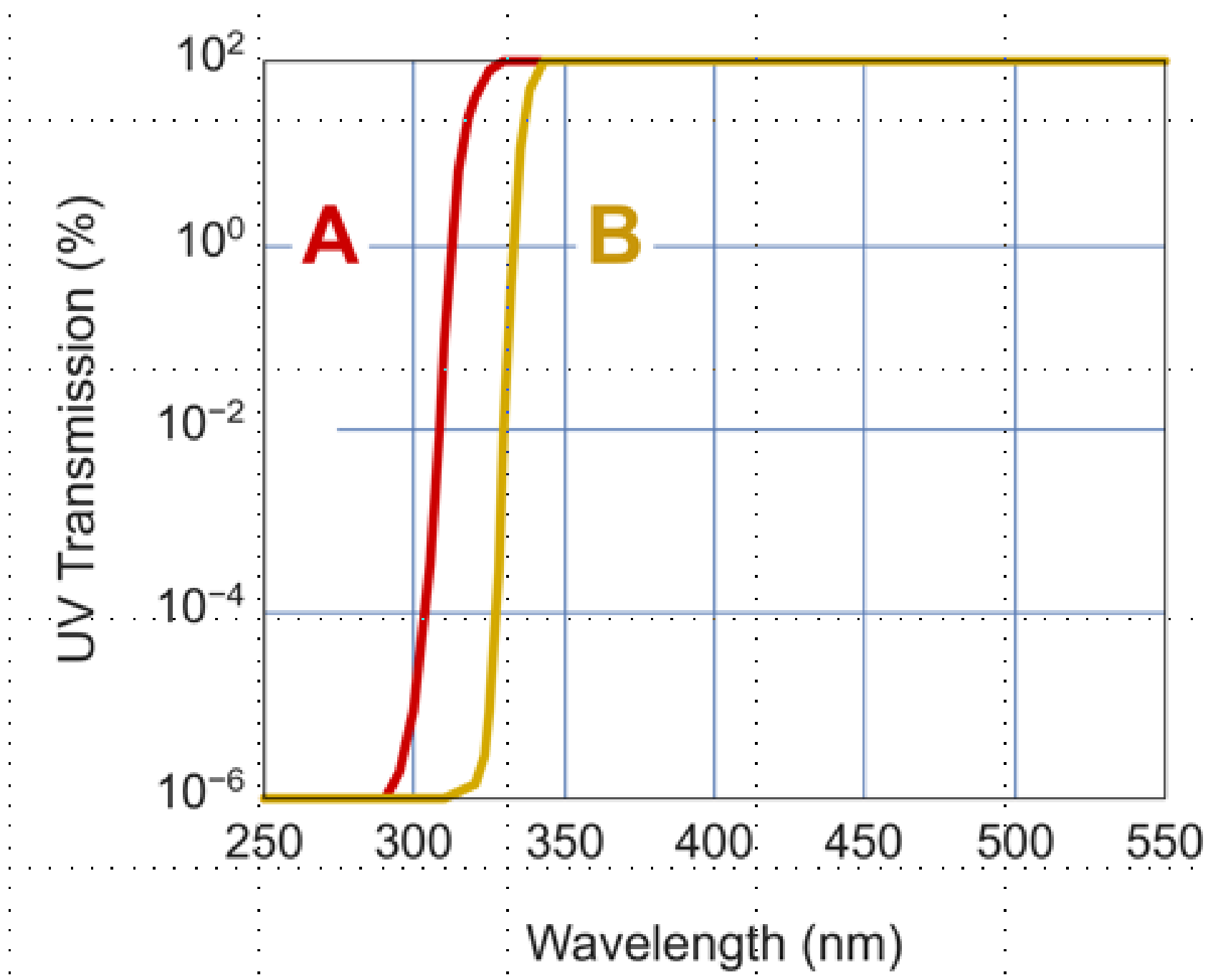

Fig 9. UV-Vis transmission characteristic of laser-blocking long-pass AlGaN filters having composition A for the 280 nm laser, and composition B for the 310 nm laser.

These are quite favorable curves because the predicted long-pass transmission is very high, while the laser light is rejected by six orders of magnitude. In the eight applications described above, the chip-user selects either filter A or filter B as appropriate for the particular laser pump employed in the application.

*D. Layout of Optoelectronic Fluoro-Chip*

If we look downward at the top face of the chip, we can gauge the positions of the LD and PD, and their areas. This is indicated schematically in Fig. 10, where we show the proposed layout for sample pumping and fluorescence capture. The input face of the PD here is covered with the laser-blocking filter of Fig. 9.

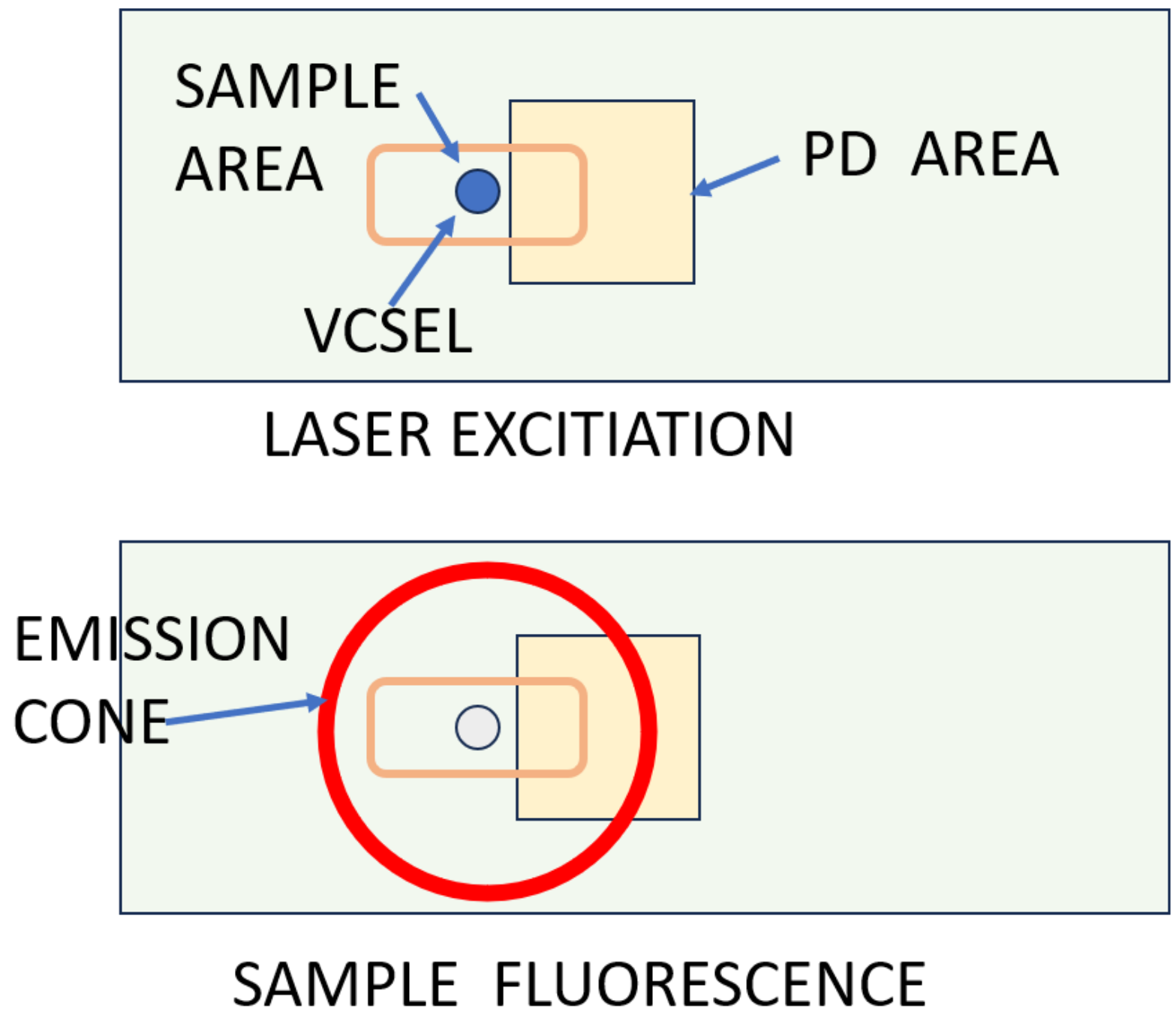

Fig. 10. Schematic top views of transparent chip showing relative size of laser pump and rectangular PD. Red circle indicates cone of fluorescent emission. The PD has a laser-blocking filter on its face.

Going into more detail, Fig. 11 presents a cross-sectional side view showing the sample contacting the top face, and the photonic-and-electronic dies bonded to the bottom face. The VCSEL cone here excites a ~spherical sample volume, ~0.5 mm diameter, with half of the resulting 4π-steradian fluorescence traveling downward, and with that light radiating into 2π steradians. Thereby, the "large area" longpass-filtered PD captures a practical fraction of that radiation, and the filter prevents any laser capture. Now we shall give detail showing that the Fig.-11 sensor is a complete self-contained lab chip.

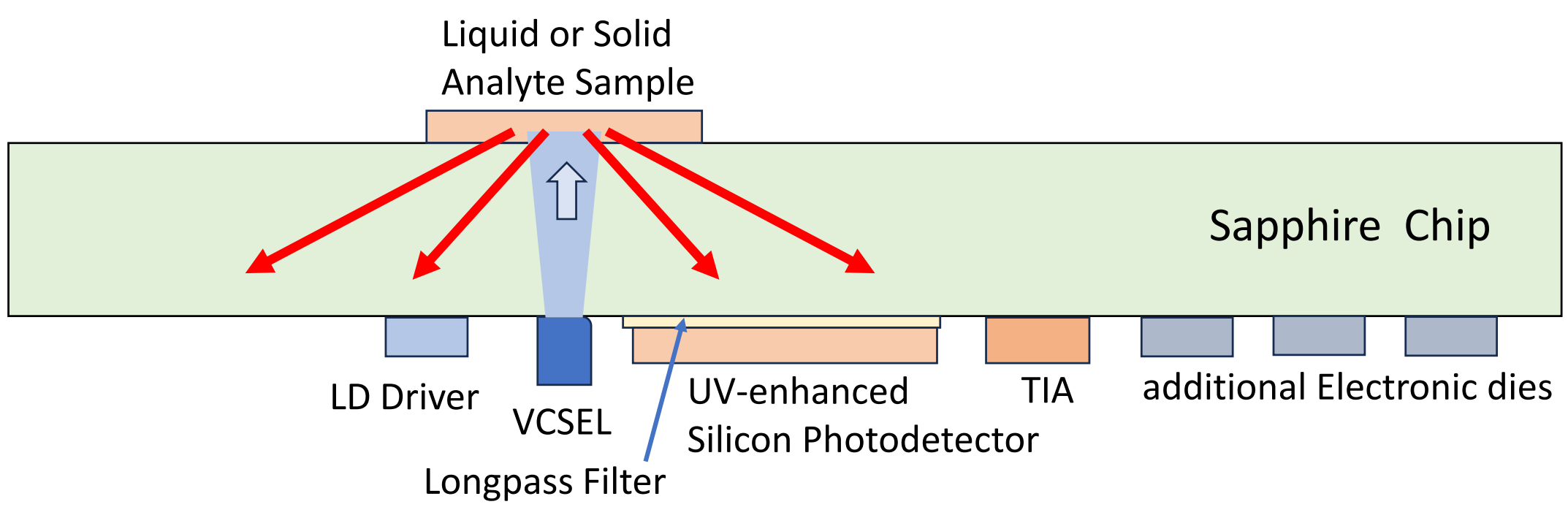

Fig. 11. Schematic cross section of chip showing size of excitation and emission zones, as well as layout of photonic and electronic chip.

*E.WiFi Battery-Powered OE Chip*

The WiFi proposal here is to use the miniature, rechargeable, Lithium-ion Polymer (LiPo) battery to power the lab chip, and to use the well-known ESP 32 electronic internet-of-things die to provide micro-processing of the PD readout as well as WiFi 802.11 and Bluetooth Low Energy (BLE) in a single chiplet. The WiFi-lab's low-cost and capable wireless fluoro-reporting benefits all eight of the above applications. The on-chip ESP32 runs two protocols simultaneously; the MQTT messaging protocol and BLE. Thus, one chip with one antenna, gives two completely different communication relationships operating in parallel. These are synergistic and give practical value to the chip. For example, using Bluetooth, a nurse or field worker could configure the OE device from a phone without touching a laptop. At the same time, MQTT over WiFi delivers chip readings directly to a personal computer (PC), cloud dashboard, or hospital information system. The battery removes a physical "tether." WiFi users could include a reviewer, a clinician, a field engineer, or a patient's wrist-- without requiring any infrastructure beyond a WiFi access point.

The technical explanation of MQTT is that ESP32 uses its WiFi radio to connect to the user's local network, then uses the TCP/IP addressing system to find a broker server, then uses MQTT formatting to publish a small message containing the fluorescence reading. The broker immediately forwards that message to any subscribed PC or dashboard. The entire round trip takes less than 0.1 seconds.

Here are the chip specifics. Figure 12 presents a bottom view of the proposed lab chip. Along the chip centerline, the PD feeds the trans-impedance amplifier (TIA), whose output is digitized by the analog to digital converter (ADC) that then

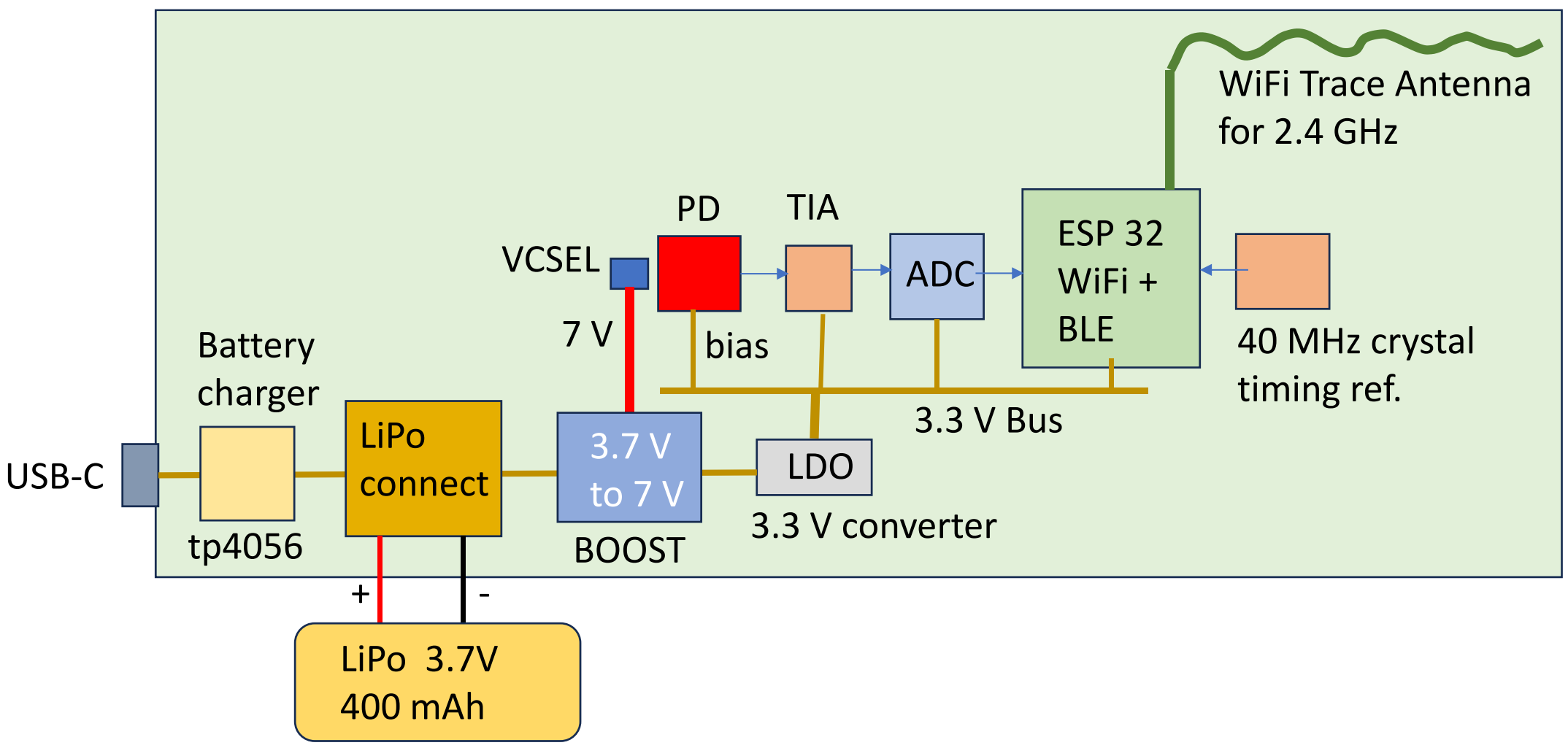

Fig.12. Schematic bottom view of WiFi OE lab chip.

Feeds the ESP32 IC. The Si PD is operated in the reverse-bias (photoconductive) mode. The PD cathode is connected to the 3.3 V rail and the PD anode is connected to the inverting input of the TIA op-amp, which sits at virtual ground. This reverse bias of ~3.3 V across the junction reduces the junction capacitance, which directly improves the TIA bandwidth and lowers the noise floor. The photocurrent generated by incident fluorescence photons flows from the junction into the TIA, which converts it to a voltage. The TIA feeds an analog-to-digital converter whose output feeds the ESP32. A small on-chip quartz crystal provides timing reference to the ESP 32. A 50-Ohm RF lead from the ESP-die contacts the on-chip microwave antenna.

Regarding electric powering, the power “zone” in the lower portion of Fig. 12 consists of the USB-C charge port, TP4056 LiPo charger IC, LiPo connector, and a separate 400 mAh LiPo cell, together with a 3.7V-to-7V Boost-converter IC driving the VCSEL. A boost-connected linear drop-out (LDO) voltage converter IC drives the electronic digital circuits.

From the LDO, there is a 3.3 V rail illustrated by gold lines, Then, Ti/Au traces run from the rail to supply PD-bias, TIA, ADC, and wireless chip. There is a 7 V Ti/Au rail running from the boost converter to the VCSEL anode. This 7V is the estimated higher-voltage needed to forward-bias the AlGaN VCSEL. Not shown in Fig. 12 is the ground return, a third metal trace that connects the negative terminal of the LiPo connector, the ground pins of every IC, the VCSEL cathode, and the PD cathode. Regarding the high-rate battery and USB connector, it will probably be necessary, for mechanical ruggedness to attach those items in a small rectangular housing (similar in size to the chip) bonded to chip bottom, such as an 18-mm thick, 24 mm x 54 mm enclosure.

### *F.Ethernet Battery-Powered OE Chip*

Ethernet chips will, for example, be deployed in hospital diagnostic labs, pharmaceutical manufacturing suites, and research institutions-- environments where clinical-grade instruments connect via wired Ethernet rather than WiFi because wired connections are deterministic, encrypted at the network switch level, and free from the radio-frequency interference that exists in every hospital building. An on-chip W5500 ethernet chiplet implements the full TCP/IP stack in hardware, not software. Fig. 13 presents the complete chip layout.

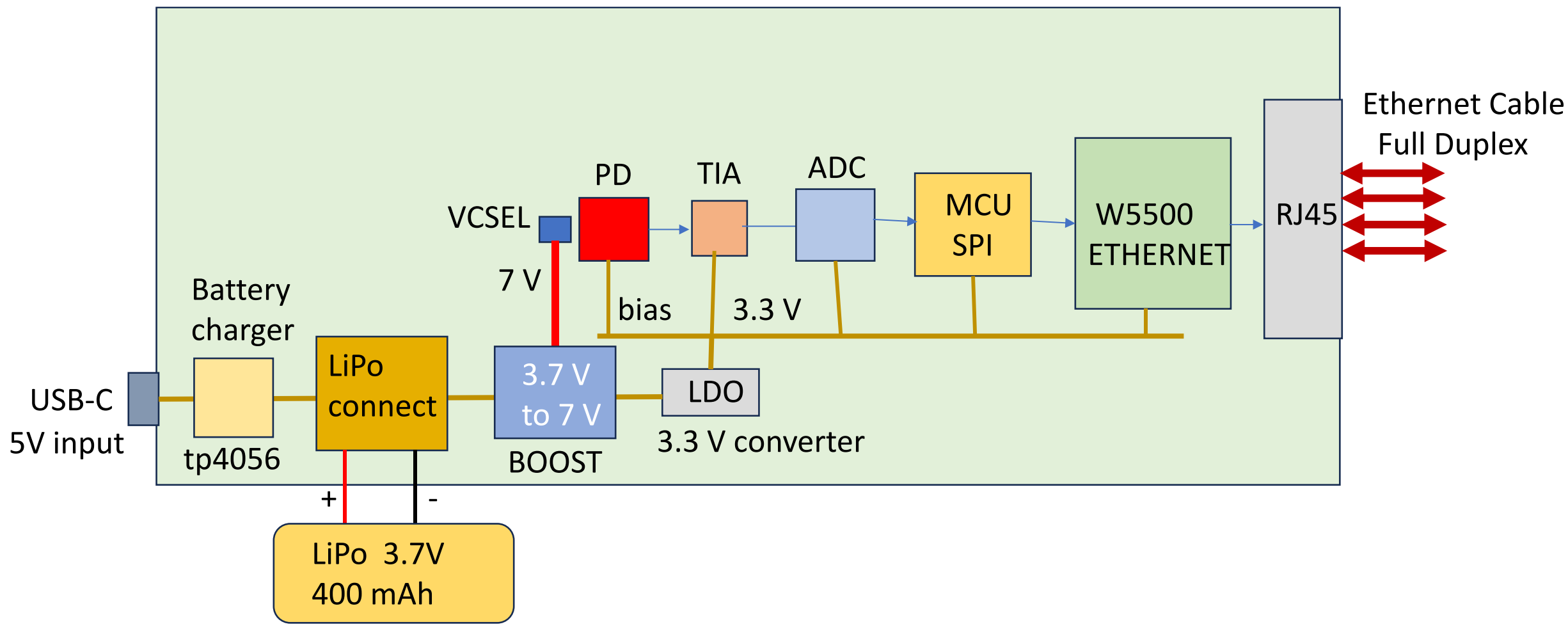

Fig.13. Schematic bottom view of Ethernet OE lab chip.

There are similarities to Fig. 12. The power zone is the same, and the 3.3 V and 7 V rails are the same. The fluoro readout begins with the same PD + TIA + ADC chiplets, but here the ADC feeds a micro-processing unit (MCU), whose output drives the ethernet chiplet.

The W5500 handles packet framing, error detection, IP addressing, TCP and UDP-- all of the rules of Ethernet and the internet protocols above it --without burdening the MCU with any of that work. It communicates with the MCU via a simple SPI bus (four wires at 3.3 V). The W5500 outputs the actual Ethernet differential signal pairs (typically at ±1V levels through internal magnetics). A magnetic transformer integrated inside of the RJ45 connector provides electrical isolation between the W5500's internal circuitry and the external cable. This is mandatory in the Ethernet standard for safety and to suppress common-mode noise. RJ 45 is the physical socket where the patch cable plugs in. Four twisted wire-pairs comprise the cable. The cable carries the Ethernet signals onwards to a network switch or PC.

## IV. SIGNAL-TO-NOISE RATIO IN APPLICATIONS

The SNRs calculated in this Section assume CW operation of the laser. However, in practice, the laser will be operated in a quasi-CW mode because of a thermal-management problem. During CW, the temperature of the laser steadily increases and this ΔT spreads to the chip. To solve this issue, the CW laser is turned on and the fluoro-measurement proceeds for 1 to 5 seconds. Then, the laser is turned off, and the user waits 4 to 5 minutes for the laser to cool, waiting for the laser temperature to return to room temperature. Then that cycle can then be repeated.

The 140-μm-diam VCSEL in Figs. 12 and 13 above has a forward-bias lasing current threshold of 2.3 A. The high-rate 10C 400 mAh battery in Figs. 12 and 13 can supply as much 4.0 A to the laser, and at 3.7A bias, the total emitted laser power will be 1 mW; the pumping power reaching the sample. We shall assume 1 mW CW VCSEL emission in these SNRs.

For each of the eight applications, the SNR was quantified by quantifying the signal properties of the lab chip, and then the noise properties of the chip were quantified. The term "signal" refers to **P_signal** which is defined as the optical signal power of the fluorescence emission collected by the PD. **P_signal** was determined here by estimating the size of four factors; **ε ($M^{-1}cm^{-1}$)**--the Molar absorptivity which is he intrinsic strength of light absorption by the fluorophore at the excitation wavelength, per molar concentration per cm of path length; **OD** – the Optical density which is the dimensionless Beer-Lambert exponent such that OD 1.0 means 90% of excitation light is absorbed and OD 0.1 means ~21%; **Abs %** -- the percentage of excitation photons actually captured by the sample, and **QY** –the Quantum yield: which is the probability (0 to 1) that an absorbed photon is re-emitted as a fluorescence photon rather than lost as heat. These quantities have been estimated for each of the eight applications, in Table 1, and from those estimates, the corresponding **P_signals** have been found as listed in Table1.

| App | Sensing Target / Conditions | λexc | Conc. | ε ($M^{-1}cm^{-1}$) | OD | Abs % | QY | P_signal |
|---|---|---|---|---|---|---|---|---|
| A | AGE / Collagen crosslinks (skin dermis, 300 µm depth) | 310 nm | 0.5 mM | 500 | 0.00750 | 1.71% | 0.07 | **25.0 nW** |
| B | Tryptophan — blood plasma (70 µM free Trp, 1 mm well) | 280 nm | 70 µM | 5 500 | 0.03850 | 8.48% | 0.13 | **230 nW** |
| C | PAH — naphthalene in water (10 ppb = 78 nM, 1 mm cell) | 280 nm | 78 nM | 7 000 | 5.46e−5 | 0.013% | 0.23 | **0.60 nW** |
| D | Bacterial TLF — water quality (50 ppb TLF = 245 nM Trp) | 280 nm | 245 nM | 5 500 | 1.35e−4 | 0.031% | 0.13 | **0.84 nW** |
| E | Ferulic acid — plant leaves (1 mM, 300 µm leaf depth) | 310 nm | 1.0 mM | 15 000 | 0.4500 | 64.5% | 0.15 | **2 086 nW** |
| F | IgG tryptophan — biopharma (1 mg/mL = 107 µM Trp eff.) | 280 nm | 107 µM | 5 500 | 0.05885 | 12.7% | 0.11 | **290 nW** |
| G | Urinary Trp — CKD monitoring (CKD: 50 µM; normal: 200 µM) | 280 nm | 50–200 µM | 5 500 | 0.0110 | 2.50% | 0.13 | **166–606 nW** |
| H | NADH — organ transplant (0.5 mM ischemic tissue, 300 µm) | 310 nm | 0.5 mM | 3 400 | 0.05100 | 11.1% | 0.019 | **45.4 nW** |

Table 1. Received Signal Power at the Photodetector.

Turning now to the noise properties, here the TIA noise floor is determined by several factors, including **R (A/W)** --the Responsivity of the S1336-33BQ photodetector: the photocurrent generated per watt of incident optical power at the emission wavelength; **$R_f$** --the transimpedance amplifier feedback resistor that converts photocurrent to output voltage; **I_signal** –the photocurrent produced by the fluorescence signal which is P_signal × R; the electrical signal the TIA must detect, in picoamperes; **V_TIA** –the TIA output voltage which is I_signal × $R_f$; the actual voltage swing available at the amplifier output, in millivolts; ; **i_Johnson** –the Johnson (thermal) noise current of the feedback resistor which is √(4kT/Rf) at BW = 1 Hz; the dominant noise source for Apps A–G; **i_shot,sig** –the Shot noise current from the signal photocurrent which is √(2q · I_signal); becomes significant only when I_signal is large as in App H; **i_total** –the total RMS noise current which is √(i²_Johnson + i²_shot,sig + i²_dark + i²_amp); the quadrature sum of all noise contributions; **SNR** –the Signal-to-noise ratio which is **I_signal / i_total** .These factors are quantified in Table 2.

| App | Target | R (A/W) | Rf | I_signal | V_TIA | i_Johnson | i_shot,sig | i_total | SNR | Dom. noise |
|---|---|---|---|---|---|---|---|---|---|---|
| A | AGE / Skin em ≈ 400 nm | 0.14 | 1 MΩ | 3 494 pA | 3.49 mV | 129 fA | 33 fA | 133 fA | **26 200×** | Johnson |
| B | Blood Plasma Trp em ≈ 340 nm | 0.10 | 1 MΩ | 22 965 pA | 23.0 mV | 129 fA | 86 fA | 155 fA | **148 000×** | Johnson |
| C | PAH Water em ≈ 340 nm | 0.10 | 100 MΩ | 60.2 pA | 6.02 mV | 12.9 fA | 4.4 fA | 15.6 fA | **3 870×** | Johnson |
| D | Bacterial TLF em ≈ 340 nm | 0.10 | 100 MΩ | 84.0 pA | 8.40 mV | 12.9 fA | 5.2 fA | 15.8 fA | **5 320×** | Johnson |
| E | Plant Phenolics em ≈ 400 nm | 0.14 | 100 kΩ | 292 040 pA | 29.2 mV | 407 fA | 306 fA | 509 fA | **574 000×** | Johnson |
| F | IgG Biopharma em ≈ 340 nm | 0.10 | 1 MΩ | 29 000 pA | 29.0 mV | 129 fA | 96 fA | 161 fA | **180 000×** | Johnson |
| G | Urinary Trp / CKD em ≈ 340 nm | 0.10 | 1 MΩ | 16 600 pA | 16.6 mV | 129 fA | 73 fA | 148 fA | **112 000×** | Johnson |
| H | NADH / Transplant em ≈ 460 nm | 0.16 | 10 MΩ | 7 261 pA | 72.6 mV | 40.7 fA | 48.2 fA | 63.5 fA | **114 000×** | Shot |

Table 2. TIA Noise Floor and Signal-to-Noise Ratio assuming a bandwidth of 1 Hz.

In sum, the SNRs predicted for applications A, B, E, F, G, H, which range from 26,200 to 574,000, are high or even extremely high. The SNRs for the "water applications" C and D are, by comparison, much weaker; however, at their SNRs of 3,870 and 5,320, respectively, these ratios seem adequate for practical use.

## V. SUMMARY

A compact, portable, wearable, micro-battery-powered lab-on-a-chip for local or network use has been proposed and analyzed. Medical or chemical or biological analytes are placed on the top surface of a UV-Vis transparent sapphire, and those molecules are excited by the UV-B/UV-C beam of an AlGaN VCSEL bonded to the sapphire bottom surface. The resulting UV-Vis fluorescence is captured and readout by a filtered, UV-enhanced Si photodetector bonded adjacent to the laser. This signal reveals the presence of targeted analytes. A set of electronic chiplets is also bonded to the chip's bottom face--synergistic circuits that enable high-performance WiFi or Ethernet networking of these lab chips.

## ACKNOWLEDGEMENT

The author acknowledges the use of Claude AI (Anthropic) as a research assistance tool during preparation of this manuscript. Claude contributed to the identification of new sensing applications, quantitative signal-to-noise calculations, electronic chiplet layout design, reference verification, and technical error checking. All scientific conclusions, judgements, and final content are the sole responsibility of the author.